\documentclass
{elsart1p}

\usepackage{times}

\usepackage{graphicx}

\usepackage{amsmath,amssymb}

\renewcommand{\vec}[1]{\boldsymbol{#1}}      
\newcommand{\ii}{\mkern 1mu\mathrm{i}\mkern 1mu}
\renewcommand{\geq}{\geqslant}
\renewcommand{\leq}{\leqslant}
\newcommand{\eqhyp}[1]{#1\nobreak\discretionary{}{\hbox{\ensuremath{#1}}}{}}
\allowdisplaybreaks

\begin{document}
\parindent 1em

\begin{frontmatter}
\vspace{-30pt}


\title{Interwall conductance in double-walled armchair carbon nanotubes}


\author[BSU]{N.A. Poklonski\corauthref{cor}},
\corauth[cor]{Corresponding author.}
\ead{poklonski@bsu.by}
\author[BSU]{Nguyen Ngoc Hieu},
\author[BSU]{E.F. Kislyakov},
\author[BSU]{S.A. Vyrko},
\author[BSU]{A.I. Siahlo},
\author[ISAN]{A.M. Popov},
\author[ISAN]{Yu.E. Lozovik\corauthref{cor}}
\ead{lozovik@isan.troitsk.ru}

\address[BSU]{Physics Department, Belarusian State University,
Minsk 220030, Belarus}
\address[ISAN]{Institute of Spectroscopy, Troitsk 142190, Moscow Region, Russia}

\begin{abstract}
The dependence of the interwall conductance on distance between walls and relative positions of walls are calculated at the low voltage by Bardeen method for $(n,n)$@$(2n,2n)$ double-walled carbon nanotubes (DWCNTs) with $n = 5, 6, \ldots, 10$. The calculations show that interwall conductance does not depend on temperature (for $T \leq 500$~K) and current-voltage characteristic is linear. The conductance decreases by 6 orders of magnitude when the interwall distance is doubled. Thus, depending on the interwall distance, DWCNTs can be used as temperature stable nanoresistors or nanocapacitors.

\end{abstract}


\end{frontmatter}

\hyphenation{nano-tube nano-tubes}
\section{Introduction}
Carbon nanotubes (CNTs) have attracted much attention from the viewpoints of basic science and technological applications because of their unique electronic and mechanical properties~\cite{Dresselhaus01}. A set of nanoelectromechanical systems (NEMS) based on relative motion of carbon nanotube walls in the multi-walled CNTs have been proposed (see Ref.~\cite{Lozovik07} for a review). Some of these NEMS, such as variable nanoresistors~\cite{Lozovik04, Yan06}, are based on tunneling current between CNT walls. The operation of these nanoresistors have been considered on example of double-walled carbon nanotubes (DWCNTs). DWCNTs produced by high-temperature treatment of single-walled nanotubes with fullerenes inside~\cite{Smith00, Bandow01}, chemical vapor deposition~\cite{Zhu02, Ren02} or arc discharge in hydrogen~\cite{Hutchison01, Saito03} have interwall distances ranging from 0.33 to 0.42 nm. However the previous calculations of tunneling current between DWCNT walls were done only for DWCNTs with interwall distance equal to 0.34 nm~\cite{Yan06, Kim02}. 

The another type of nanotube-based NEMS is a nanocapacitor with coaxial nanotube walls used as its plates (for example, nanotube-based variable nanocapacitor can be used for controlling of operation of three-terminal memory cell by electrostatic forces~\cite{Popov07}). In such a nanocapacitor the interwall tunneling current is a leakage current and interwall distance should be chosen so that tunneling current is negligible. Thus calculations of interwall tunneling current as a function of interwall distance is an actual problem for elaboration of nanotube-based NEMS.

In this Letter, we calculate the dependence of the interwall conductance in DWCNTs with armchair walls on distance between nanotube walls at the low voltage using the Bardeen formalism within non-orthogonal tight-binding approximation. The influence of temperature on the interwall conductance is investigated. According to semiempirical calculations~\cite{Kolmogorov00, Saito01, Belikov04} and calculations based on density functional theory~\cite{Bichoutskaia06} the interwall interaction energy of DWCNTs with armchair walls depends on the relative positions of walls. Here we show that the total transmission and therefore conductance of DWCNTs with armchair walls depends also on the relative positions of walls.

\section{Tunneling matrix element in DWCNTs}

In the tight-binding approximation, the wave function of graphene can be expressed as~\cite{Barnett05}
\[
   \Psi(\vec{k},\vec{r}) = \frac{1}{\sqrt{N_\text{G}}} \sum_{g=1}^{N_\text{G}}\exp(\ii\vec{k}\cdot\vec{R}_g) \frac{1}{\sqrt{2}} \biggl(\chi(\vec{r} - \vec{R}_g) \pm \frac{\omega(\vec{k})}{|\omega(\vec{k})|} \chi(\vec{r} - \vec{R}_g - \vec{d})\biggr),
\]
where $\vec{k}$ is a two-dimensional vector in reciprocal space of graphene lattice with lattice constant $a = \sqrt{3}a_\text{C--C}$ ($a_\text{C--C}$ is the C--C bond length), $\vec{d} = (\vec{a}_1 + \vec{a}_2)/3$ is the vector between two atoms in the unit cell, $\omega(\vec{k}) = 1 + \exp(-\ii\vec{k}\vec{a}_1) + \exp(-\ii\vec{k}\vec{a}_2)$, $\vec{R}_g = g_1\vec{a}_1 + g_2\vec{a}_2$, here $\vec{a}_1$ and $\vec{a}_2$ are the lattice vectors of graphene, and $N_\text{G}$ is the number of the graphene unit cells. The function $\chi(\vec{r})$ is the Slater $2p_x$ orbital~\cite{Clementi63}. The signs $\pm$ correspond to the $\pi$ and $\pi^*$ configurations of orbitals in graphene.

For $(n,n)$@$(2n,2n)$ DWCNTs rotational period of the inner wall $2\pi/n$ equals to 2 rotational periods of the external wall, and their translational periods are coincide and equal a. Therefore the structure of the DWCNT is determined by the structure of the primitive unit cell restricted by cylindrical coordinates $0 \leq \varphi < 2\pi/n$, $0 \leq z < a$. These DWCNTs are commensurate and belong to the ``strong'' coupling case in the terminology of~\cite{Tunney06}. 

\begin{figure}
\hfil\includegraphics{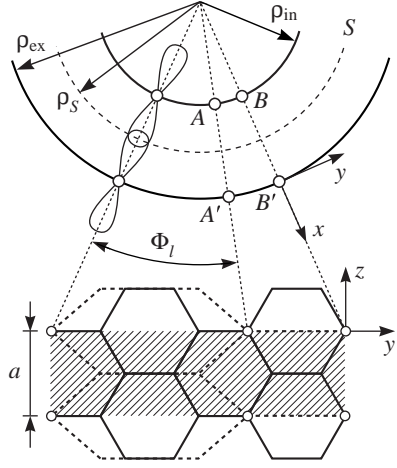} 
\caption{
Geometry of the $(n,n)$@$(2n,2n)$ DWCNT corresponding to the relative position of the walls with $\Delta\Phi = 0$ and $\Delta Z = 0$. Primitive unit cell (shaded area) is represented in the lower part in which the dashed line corresponds to the inner wall projection on the external one (solid line). The number of coincidental atoms in the primitive unit cell is 2 and in the DWCNT translational unit cell is $n_\text{c} = 2n$. The value $n_\text{c} = 2n$ repeats with periods $\varphi_\text{c} = \pi/10$ and $a_\text{c} = a/2$ at the relative rotation of walls and their relative displacement along the DWCNT axis (in $z$ direction), respectively.}\label{fig:01}
\end{figure}

An 1D Bloch function of $(n,n)$ CNT in state $m$ for a $2p_x$ orbital $\chi_{A,B}$ ($A$ and $B$ in Fig.~\ref{fig:01} correspond to two nonequivalent atoms of graphene unit cell) in cylindrical coordinates $(\rho,\varphi,z)$ is (see Ref.~\cite{Saito93})
\[
   \Psi_{A,B}^m(k;\rho,\varphi,z) = \frac{1}{\sqrt{nN}} \sum_{j=1}^N \exp(\ii kZ_j) \sum_{l=1}^n \exp(\ii m\Phi_l)\chi_{A,B}(\rho - \rho_\text{w}, \varphi - \Phi_l, z - Z_j),
\]
where $k$ is the wavenumber of an electron along CNT, $m = 0, 1, \ldots, n - 1$ is the subband number, $Z_j$ and $\Phi_l = 2\pi l/n$ are cylindrical coordinates of carbon atoms, $\rho_\text{w}$ is the radius of the CNT, and $N$ is the number of nanotube translational unit cells. The second sum corresponds to rotational symmetry of the CNT.

In the Bardeen formalism~\cite{Bardeen61}, the tunneling matrix element between states $\Psi_\text{in}$ and $\Psi_\text{ex}$ of the inner (in) and the external (ex) walls of a DWCNT, respectively, is
\[
   M_\text{in,ex} = \frac{\hbar^2}{2m_0} \int_S (\Psi_\text{in}^*\nabla\Psi_\text{ex} - \Psi_\text{ex}\nabla\Psi_\text{in}^*)\cdot \mathrm{d}\vec{S},
\]
where $S$ is an arbitrary cylindrical surface with radius $\rho_S$ (see Fig.~\ref{fig:01}) between the inner and the external walls, $\hbar$ is the Planck constant, and $m_0$ is a mass of the electron in vacuum.

The matrix element of the probability density current operator $\vec{J}$ between states $\Psi_\text{in}$ and $\Psi_\text{ex}$ of the inner and the external walls for $\chi_\text{in}$ and $\chi_\text{ex}$ orbitals is
\begin{align*}
   M_\text{in,ex}^{m_\text{in}m_\text{ex}} &= \langle\Psi_\text{ex}^{m_\text{ex}}(k_\text{ex})|\vec{J}|\Psi_\text{in}^{m_\text{in}}(k_\text{in})\rangle \\
   &= \frac{1}{N\sqrt{n_\text{ex}n_\text{in}}} \sum_{j = 1}^N \sum_{j'=1}^N \exp[\ii(k_\text{in}Z_{j} - k_\text{ex}Z_{j'})] \sum_{l'=1}^{n_\text{ex}} \sum_{l=1}^{n_\text{in}} \exp[\ii(m_\text{in}\Phi_l - m_\text{ex}\Phi_{l'})] \\
   &\times\langle\chi_\text{ex}(\rho - \rho_\text{ex}, \varphi - \Phi_{l'}, z - Z_{j'})|\vec{J}|\chi_\text{in}(\rho - \rho_\text{in}, \varphi - \Phi_l, z - Z_j)\rangle,
\end{align*}
where $\rho_\text{in}$ and $\rho_\text{ex}$ (see Fig.~\ref{fig:01}) are the radii of the inner (in) and the external (ex) walls, respectively.

Figure~\ref{fig:01} shows that for $(n,n)$@$(2n,2n)$ DWCNTs in the case where one or two atoms of the primitive unit cell of the inner wall have the same cylindrical coordinates with some atoms of the outer wall $\Phi_{l'} = \Phi_l$ and $Z_{j'} = Z_j$, the differences $|\Phi_{l'} - \Phi_l|$ and $|Z_{j'} - Z_j|$ for other pairs of atoms are essential and their overlaps are small. Therefore only the interaction between $2p_x$ orbitals of atoms with the same cylindrical coordinates $\Phi_{l'} = \Phi_l$ and $Z_{j'} = Z_j$ is considered. In this case the matrix element can be rewritten as
\begin{align}\label{eq:01}
   M_\text{in,ex}^{m_\text{in}m_\text{ex}} &= \frac{1}{n_\text{in}N} \sum_{j=1}^N \exp[\ii(k_\text{in} - k_\text{ex})Z_j)] \sum_{l=1}^{n_\text{in}} \exp[\ii(m_\text{in} - m_\text{ex})\Phi_l)] \notag\\
   &\times\langle\chi_\text{ex}(\rho - \rho_\text{ex}, \varphi - \Phi_l, z - Z_j)|\vec{J}|\chi_\text{in}(\rho - \rho_\text{in}, \varphi - \Phi_l, z - Z_j)\rangle \notag\\
   &= \delta_{k_\text{in},k_\text{ex}}\delta_{m_\text{in},m_\text{ex}} \frac{\hbar^2}{2m_0} \int_S \biggl(\chi_\text{in}^* \frac{\mathrm{d}}{\mathrm{d}\rho}\chi_\text{ex} - \chi_\text{ex} \frac{\mathrm{d}}{\mathrm{d}\rho} \chi_\text{in}^*\biggr)\, \mathrm{d}S,
\end{align}
where $\delta_{k_\text{in},k_\text{ex}}$ and $\delta_{m_\text{in},m_\text{ex}}$ are Kronecker symbols ensuring impulse and $z$ projection of orbital momentum conservations in the electron tunneling process, respectively.

We take $\rho_S = (\rho_\text{in} + \rho_\text{ex})/2$. In this case $(\mathrm{d}\chi_\text{ex}/\mathrm{d}\rho)_{\rho = \rho_S} = -(\mathrm{d}\chi_\text{ex}/\mathrm{d}\rho)_{\rho = \rho_S}$, and the matrix element is
\begin{equation}\label{eq:02}
   M_\text{in,ex}^{m_\text{in}m_\text{ex}} = \delta_{k_\text{in},k_\text{ex}}\delta_{m_\text{in},m_\text{ex}} \frac{\hbar^2}{m_0} \int_S \chi_\text{in}^* \frac{\mathrm{d}}{\mathrm{d}\rho} \chi_\text{ex}\,\mathrm{d}S,
\end{equation}
where $\chi = (\xi^5/32\pi)^{1/2} \cfrac{\rho_\text{ex} - \rho_\text{in}}{2} \exp(-\xi r/2)$ is the Slater orbital; $\xi = \text{1.5679}/a_\text{B}$~\cite{Clementi63}, with $a_\text{B}$ is the Bohr radius, and $r$ is module of the radius vector from atom centre.

\begin{figure}
\hfil\includegraphics{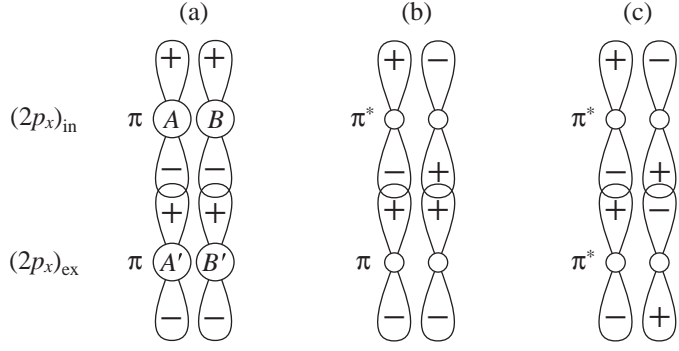} 
\caption{
Configuration of interaction between two states $\pi$ and $\pi^*$ for atoms with the same cylindrical coordinates $\Phi$ and $Z$ in DWCNT.}\label{fig:02}
\end{figure}

The configurations of interaction between two states $\pi$ and $\pi^*$ for the atoms with the same cylindrical coordinates $\Phi$ and $Z$ in DWCNT are shown in Fig.~\ref{fig:02}. In the case of $\pi\pi^*$ interaction (Fig.~\ref{fig:02}(b)) the matrix element is zero because of opposite signs for $A$ and $B$ interacting orbitals.

\section{Interwall conductance in double-walled armchair carbon nanotubes}
When a voltage $U$ is applied between the inner (in) and the external (ex) walls, the tunneling current $I$ can be written as~\cite{Tersoff85}
\begin{equation}\label{eq:03}
   I = \frac{2\pi e}{\hbar} \sum_\text{in} \sum_\text{ex} [f(E_\text{in}) - f(E_\text{ex} + eU)]|M_\text{in,ex}^{m_\text{in}m_\text{ex}}|^2\delta(E_\text{in} - E_\text{ex}),
\end{equation}
where $e$ is the electron charge module, and $f(E_\text{in})$ and $f(E_\text{ex})$ are the Fermi--Dirac functions. The delta function $\delta(E_\text{in} - E_\text{ex})$ ensures the energy conservation for the tunneling between walls. The summation is carried out over $k$ and $m$ quantum numbers of walls.

\begin{figure}
\hfil\includegraphics{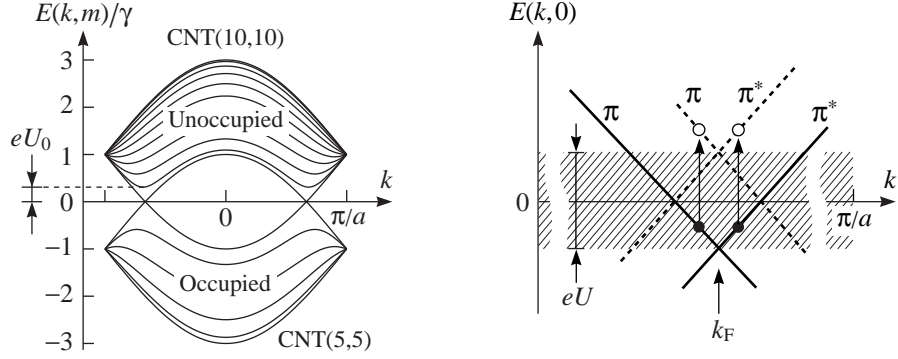} 
\caption{
Schematic band structure of the (5,5)@(10,10) DWCNT in the equilibrium state (left) and under bias $|U| < U_0$ (enlarged); $k_\text{F} = 2\pi/3a$ and $E_\text{F} = 0$.}\label{fig:03}
\end{figure}

The energy dispersion relations $E(k,m)$ for the armchair $(n,n)$ CNTs are written as~\cite{Dresselhaus01}
\begin{equation}\label{eq:04}
   E(k,m) = \pm\gamma \sqrt{1 \pm 4\cos\biggl(\frac{m\pi}{n}\biggr) \cos\biggl(\frac{ka}{2}\biggr) + 4\cos^2\biggl(\frac{ka}{2}\biggr)},
\end{equation}
where $\gamma = 2.6$~eV is resonance integral~\cite{Wildoer98, Odom98}, $k$ is the wavenumber of an electron along the CNT, and $m = 0, 1, \ldots, n - 1$ is the subband number. According to Eq. \eqref{eq:04} for $(2n,2n)$ CNT the distance between $E_\text{F} = 0$ and minimum of $m = \pm1$ band is $eU_0 = \gamma\sqrt{1 - \cos^2(\pi/2n)} \approx \gamma\pi/2n$. The band structure of the (5,5)@(10,10) DWCNT is presented in Fig.~\ref{fig:03}. In this case, $eU_0 \approx 0.8$~eV and for bias $|U| < U_0$ for temperature $T \to 0$ the tunneling occurs only between states with $m_\text{in} = m_\text{ex} = 0$. In this voltage domain, the energy bands $E(k,0)$ are linear functions of $k$ and density of states is constant. Direct calculation of the $k$-integral in Eq. \eqref{eq:03} for the armchair nanotube of length $L = Na$ gives (see Appendix)
\begin{equation}\label{eq:05}
   I = \frac{2\pi e^2U}{\hbar} |M_\text{in,ex}^{00}|^2D_\text{in}(E_\text{F})D_\text{ex}(E_\text{F})L^2,
\end{equation}
where $M_\text{in,ex}^{00}$ is the tunneling matrix element for the translational unit cell in the case $m_\text{in} = m_\text{ex} = 0$, $D_\text{in}(E_\text{F})$ and $D_\text{ex}(E_\text{F})$ are the densities of states per unit length along the nanotube axis at the Fermi level $E_\text{F} = 0$ for the inner and the external walls, respectively. The formula \eqref{eq:05} shows that the tunneling current at $|U| < U_0$ does not depend on temperature when $k_\text{B}T < 0.1eU_0$ ($0.1eU_0$ for $n = 5, 6, \ldots, 10$ corresponds to 500~K) and current-voltage characteristic is linear.

Formula \eqref{eq:05} is also valid for any non-constant $D_\text{in}(E_\text{F})$ and $D_\text{ex}(E_\text{F})$ in the low temperature limit ($T \to 0$) and small voltage (compare with Ref.~\cite{Tersoff85}). For armchair $D_\text{in}(E_\text{F}) = D_\text{ex}(E_\text{F}) = 8/(\pi\gamma a\sqrt{3})$, i.e., the interwall tunneling conductance per one translational unit cell ($L = a$) is
\begin{equation}\label{eq:06}
   G = \frac{I}{U} = \frac{128e^2}{3\pi\hbar\gamma^2} |M_\text{in,ex}^{00}|^2.
\end{equation}

We have studied the influence of the relative position of walls of the DWCNT on the tunneling matrix element. This position is determined by the angle $\Delta\Phi$ of the relative rotation of walls and the relative displacement $\Delta Z$ of walls along the nanotube axis, see Fig.~\ref{fig:01}. For a discrete set of relative positions of walls ($\Delta\Phi_\text{c}, \Delta Z_\text{c}$) some atoms of the inner and the external walls have the same cylindrical coordinates $\Phi$ and $Z$. The number $n_\text{c}$ of these ``coincidental'' atoms in translational unit cell of $(n,n)$@$(2n,2n)$ DWCNT (which consists of $n$ primitive unit cells) can be written as 
\[
   n_\text{c} = \bigg\{ \renewcommand{\arraystretch}{1.1}
\begin{array}{ll}2n,&~~ \text{if $(\Delta\Phi_\text{c},\Delta Z_\text{c}) = (q\pi/10, pa/2)$, where $q$ and $p$ are integers},\\
n,&~~ \text{if $(\Delta\Phi_\text{c},\Delta Z_\text{c}) = (q\pi/30, pa/2)$, where $q$ is not multiple of 3.}\end{array}
\]

For relative positions of walls $(\Delta Z,\Delta\Phi) = (\Delta Z_\text{c},\Delta\Phi_\text{c})$, we numerically calculate the transfer matrix element for one pair of $p_x$ orbitals $\mu_\text{in,ex}^{00} = M_\text{in,ex}^{00}/n_\text{c}$ using Eq. \eqref{eq:02}. The calculations for the (5,5)@(10,10) DWCNT give $|\mu_\text{in,ex}^{00}| ={}$0.325~eV for one pair of $p_x$ orbitals. This value is approximately equal to the value used in Ref.~\cite{Saito93}. 

\begin{figure}[!b]
\centering
\begin{minipage}[b]{0.48\linewidth} 
\hfil\includegraphics{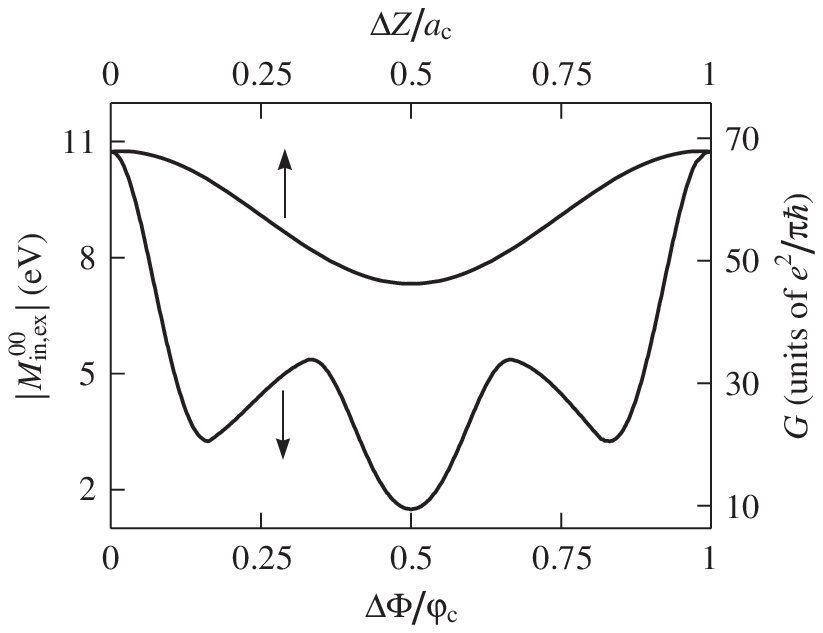} 
\caption{
Dependences of square of the tunneling matrix element $|M_\text{in,ex}^{00}|^2$ and conductance $G$ for the (5,5)@(10,10) DWCNT per translational unit cell on the relative position of walls, where $\varphi_\text{c} = \pi/10$ and $a_\text{c} = a/2$ are the angular and translational periods of these dependences for the (5,5)@(10,10) DWCNT.}\label{fig:04}
\end{minipage}
\hspace{0.5cm} 
\begin{minipage}[b]{0.48\linewidth}
\hfil\includegraphics{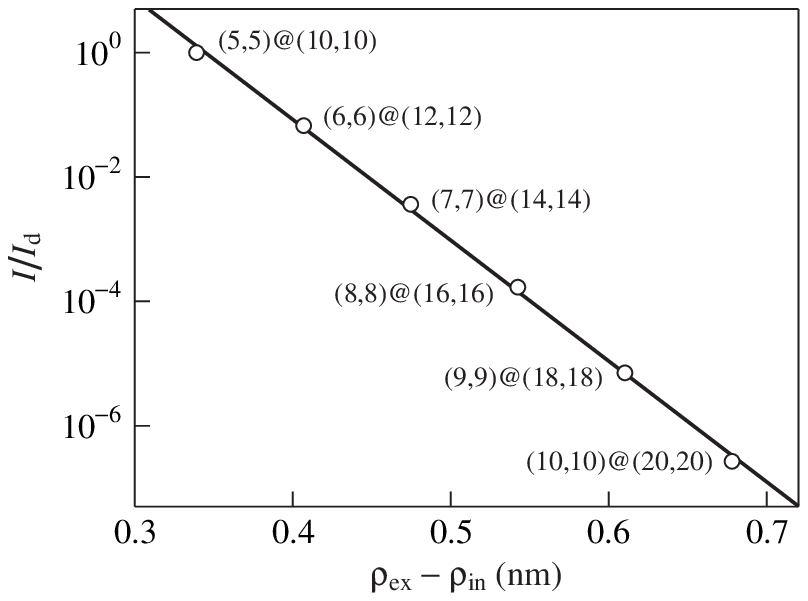} 
\caption{
Dependence of ratio of the interwall currents $I/I_\text{d}$ on the distance between walls ($\rho_\text{ex} - \rho_\text{in}$), where $I$ and $I_\text{d}$ are the interwall currents of the $(n,n)$@$(2n,2n)$ and (5,5)@(10,10) DWCNTs, respectively, $\rho_\text{in}$ and $\rho_\text{ex}$ are the radii of the inner and the external walls in a DWCNT. The straight line is the fit of values. The calculated results correspond to $\Delta\Phi = 0$ and $\Delta Z = 0$.}\label{fig:05}
\end{minipage}
\end{figure}

For small relative displacements of walls from the relative positions $(\Delta\Phi,\Delta Z) = (\Delta\Phi_\text{c},\Delta Z_\text{c})$ we neglect the ``broadening'' of $\delta$\nobreakdash-s in Eq. \eqref{eq:01} and calculate dependence of the matrix element of the (5,5)@(10,10) DWCNT on $\Delta\Phi$ and $\Delta Z$ according to Eq.~\eqref{eq:02} taking into account the dependence of $\chi$ on $\Delta\Phi$ and $\Delta Z$. Figure~\ref{fig:04} shows the dependences of square of the tunneling matrix element per one translational unit cell $|M_\text{in,ex}^{00}|^2$ and the conductance $G$ measured in quantum units $e^2/\pi\hbar$ at $\gamma = 2.6$~eV on $\Delta\Phi/\varphi_\text{c}$ and $\Delta Z/a_\text{c}$, where $\varphi_\text{c} = \pi/10$ and $a_\text{c} = a/2$ are the angular and translational periods of these dependences for the (5,5)@(10,10) DWCNT, respectively (Fig.~\ref{fig:01}).

Note, that the number of coincidental atoms (for $\Delta\Phi = 0$, $\Delta Z = 0$) per one translational unit cell in $(n,n)$@$(2n,2n)$ DWCNTs is $2n$, which must be taken into account in calculating the tunneling current in this case. Figure~\ref{fig:05} shows the calculated dependence of the interwall current $I/I_\text{d}$ for $\Delta\Phi = 0$ and $\Delta Z = 0$ on the interwall distance $\rho_\text{ex} - \rho_\text{in}$ at the low applied voltage, where $I$ and $I_\text{d}$ are the interwall currents in $(n,n)$@$(2n,2n)$ and the (5,5)@(10,10) DWCNTs, respectively, with $n = 5, 6, \ldots, 10$. It is seen that the ratio of the interwall currents of DWCNTs with armchair walls decreases nearly exponentially with increase of the distance between the nanotube walls. The line in Fig.~\ref{fig:05} corresponds to the fit $\log(I/I_\text{d}) \approx 15.4 - 1.94(\rho_\text{ex} - \rho_\text{in})$, where $\rho_\text{ex}$ and $\rho_\text{in}$ are measured in nanometers.

\section{Conclusion}
\begin{figure}
\hfil\includegraphics{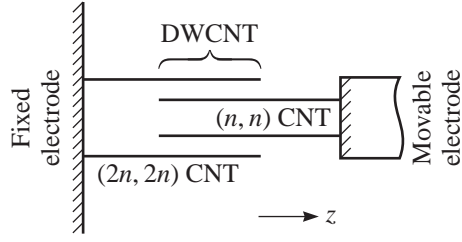} 
\caption{
Scheme of a DWCNT-based NEMS which (depending on the interwall distance) can be used as a variable nanoresistor or variable nanocapacitor.}\label{fig:06}
\end{figure}

In conclusion, the interwall conductance in the DWCNTs has been studied using the Bardeen method. Numerical calculations show that in the (5,5)@(10,10) DWCNT relative sliding of the walls along the DWCNT axis leads to 30\% variation of the interwall tunneling current, while the dependence on the angle of relative wall rotation is more pronounced. It is found that conductance of $(n,n)$@$(2n,2n)$ DWCNT for $n = 5, 6, \ldots, 10$ does not depend on temperature for $T \leq{}$500~K. The interwall conductance $G$ of $(n,n)$@$(2n,2n)$ DWCNTs decreases by 6 orders of magnitude when the interwall distance is increased from 0.34~nm to 0.68~nm. Thus DWCNTs can be used as temperature stable (up to 500~K) nanoresistors in the case of small interwall distances and as nanocapacitors with negligible leakage current in the case of interwall distances $\geq{}$1~nm. The scheme of these NEMSs is shown in Fig.~\ref{fig:06}.

\section*{Acknowledgements}
This work is supported by the Belarusian Foundation for Basic Research (Grant No. F08R-061) and Russian Foundation for Basic Research (Grants Nos. 08-02-90049-Bel and 08-02-00685).

\appendix
\section{Calculations of the tunneling current}
In the case $m_\text{in} = m_\text{ex} = 0$, from Eq. \eqref{eq:04} we obtain $E_\text{in}(k, m=0) = E_\text{ex}(k, m=0) = E_0(k) = \pm\gamma[1 - 2\cos(ka/2)]$, and the formula \eqref{eq:03} for the tunneling current in DWCNTs with the length of $L$ (using $\sum_k = (L/2\pi)\int \mathrm{d}k$) can be written as
\begin{align}\label{eq:A1}
   \frac{I(U,T)}{L^2} &= \frac{64e}{2\pi\hbar} \int_\text{BZ} \mathrm{d}k_\text{in} \int_\text{BZ} \mathrm{d}k_\text{ex} \{f[E_0(k_\text{in})] - f[E_0(k_\text{ex}) + eU]\}|M_\text{in,ex}^{00}|^2\delta[E_0(k_\text{in}) - E_0(k_\text{ex})]\delta_{k_\text{in},k_\text{ex}} \notag\\
   &=  \frac{32e}{\pi\hbar}|M_\text{in,ex}^{00}|^2 \biggl(\frac{\mathrm{d}E_0}{\mathrm{d}k_\text{in}}\biggr)^{-1}_{\!k_\text{in}=k_\text{F}} \int_\text{BZ}\{f[E_0(k_\text{in})] - f[E_0(k_\text{in}) + eU]\}\,\mathrm{d}k_\text{in} = \frac{32e}{\pi\hbar}|M_\text{in,ex}^{00}|^2 \biggl(\frac{\mathrm{d}E_0}{\mathrm{d}k_\text{in}}\biggr)^{-1}_{\!k_\text{in}=k_\text{F}} A,
\end{align}
where $\displaystyle A = \int_\text{BZ}\{f[E_0(k_\text{in})] - f[E_0(k_\text{in}) + eU]\}\,\mathrm{d}k_\text{in}$, $k_\text{F} = 2\pi/3a$, and $E_\text{F} = 0$. The integral is over the first Brillouin zone (BZ). Factor 64 stands for summing over $\pi$ and $\pi^*$ bands with $m_\text{in} = m_\text{ex} = 0$, two wave vector directions and two spin directions of an electron for both CNTs.

In order to calculate the integral $A$, we assume that $E_0(k)$ is linear in the domain of integration and use variable $\displaystyle k' = \frac{k_\text{in} - k_\text{F}}{k_\text{B}T}\eqhyp\times \frac{\mathrm{d}E_0(k_\text{in})}{\mathrm{d}k_\text{in}}\bigg|_{k_\text{in}=k_\text{F}} = (k_\text{in} - k_\text{F})\alpha$, therefore, $dk' = \alpha\,\mathrm{d}k_\text{in}$.

The integral $A$ can be rewritten as
\[
   A = \frac{1}{\alpha} \biggl(\int_{-\infty}^{+\infty} \frac{\mathrm{d}k'}{1 + \exp k'} - \int_{-\infty}^{+\infty} \frac{\mathrm{d}k'}{1 + \exp(eU/k_\text{B}T + k')}\biggr) = \frac{1}{\alpha} \biggl(\int_0^\infty \frac{\mathrm{d}\zeta}{\zeta(1 + \zeta)} - \int_0^\infty \frac{\mathrm{d}\zeta}{\zeta[1 + \zeta\exp(eU/k_\text{B}T)]}\biggr),
\]
where $\zeta = \exp(k')$.

Hence
\begin{equation}\label{eq:A2}
   A = \frac{1}{\alpha} \biggl[-\ln\biggl(\frac{1 + \zeta}{\zeta}\biggr) + \ln\biggl(\frac{1 + \zeta\exp(eU/k_\text{B}T)}{\zeta}\biggr)\biggr]\bigg|_0^\infty = eU \biggl(\frac{\mathrm{d}E_0}{\mathrm{d}k_\text{in}}\biggr)_{\!k_\text{in}=k_\text{F}}^{-1} = \frac{\pi}{4} eUD(E_\text{F}).
\end{equation}
Inserting \eqref{eq:A2} into \eqref{eq:A1}, we obtain formulae \eqref{eq:05} and \eqref{eq:06}.

\end{document}